\documentclass[pra,aps,amsmath,amssymb,superscriptaddress,twocolumn, longbibliography]{revtex4-1}
\usepackage{graphicx}
\usepackage{color,outlines}
\usepackage[english]{babel}
\usepackage{amsthm}
\usepackage[sc,osf]{mathpazo}
\usepackage{hyperref}
\usepackage{physics}
\usepackage{outlines}
\usepackage{tabularx}
\usepackage{mathtools}
\usepackage{lipsum}

\definecolor{green2}{RGB}{0,100,0}

\usepackage{tikz, pgfplots}
\usetikzlibrary{positioning}

\begin{document}
	
	\title{Krylov complexity in ergodically constrained nonintegrable transverse-field Ising model}
	
	\author{Gaurav Rudra Malik}
	\email{gauravrudramalik.rs.phy22@itbhu.ac.in}
	\affiliation{Indian Institute of Technology (Banaras Hindu University), Varanasi, India~221005} 
	
	\author{Jeet Sharma}
	\email{jeet.sharma@niser.ac.in}
	\affiliation{National Institute of Science Education and Research, Bhubaneshwar, India~752050} 
	
	\author{Rohit Kumar Shukla}
	\email{rohitkrshukla.rs.phy17@itbhu.ac.in}
	\affiliation{Department of Chemistry; Institute of Nanotechnology and Advanced Materials; Center for Quantum Entanglement Science and Technology, Bar-Ilan University, Ramat-Gan 5290002, Israel} 
	
	\author{S. Aravinda}
	\email{aravinda@iittp.ac.in}
	\affiliation{Indian Institute of Technology Tirupati, Tirupati, India~517619} 
	
	\author{Sunil Kumar Mishra}
	\email{sunilkm.app@iitbhu.ac.in}
	\affiliation{Indian Institute of Technology (Banaras Hindu University), Varanasi, India~221005} 
	
	\begin{abstract}
		
		The nonintegrable transverse-field Ising model is a common platform for studying ergodic quantum dynamics. In this work, we introduce a simple variant of the model in which this ergodic behaviour is suppressed by introducing a spatial inhomogeneity in the interaction strengths. For this we partition the chain into two equal segments within which the spins interact with different coupling strengths. The ratio of these couplings defines an inhomogeneity parameter, whose variation away from unity leads to constrained dynamics. We characterize this crossover using multiple diagnostics, such as the long-time saturation of out-of-time-ordered correlators, level-spacing statistics, and the spectral form factor. We further examine the consequences for operator growth in Krylov space and for entanglement generation in the system's eigenstates. Together, these results demonstrate that introducing a macroscopic inhomogeneity in coupling strengths provides a minimal, disorder-free route to breaking ergodicity in this specific model of interacting spins. 
		
	\end{abstract}

	\maketitle
	\section{Introduction}
	\label{Introduction}
	The dynamics of isolated quantum systems in the long-time limit remains an open problem and an active area of research in modern physics. A specific direction of inquiry focuses on the emergence of statistical mechanics following the unitary dynamics of quantum systems, known widely as the Eigenstate Thermalisation Hypothesis (ETH) \cite{Deutsch_Thermalisation, Srednicki_Thermalisation}. Our current understanding indicates that a majority of quantum systems undergoing unitary dynamics ultimately leading to the thermal state, given by the Gibbs ensemble \cite{DAlessio_ETH, nandkishore_huse_2015, Polkovnikov_RMP_2011}. This convergence also underlines phenomena such as quantum chaos and associated features such as information scrambling and entanglement generation \cite{Bohigas_BGS, Srednicki_RMT_1999, Haake_QuantumChaos}.
	
	However, while ETH defines a generic tendency of interacting quantum systems under dynamical evolution, there exist several counter examples that deviate from the notion of thermalisation \cite{Basko_MBL, Oganesyan_Huse_2007}. These are models that show ergodicity breaking mechanisms, and are often at the center of several exotic physical phenomena \cite{Rigol_GGE}. Moreover, ever since the emergence of quantum simulators making use of ultracold atoms in optical lattices \cite{Bloch_ColdAtoms, Schreiber_MBL_Experiment} and superconducting circuits \cite{Guo_SC_Ergodicity_2019, Xu_MBL_SC_2018, Nature_OTOC_2025}, it is has also become possible to experimentally observe the dynamical features of quantum models \cite{Bernien_QuantumSimulator_2017, Roushan_Scrambling_2017, Georgescu_QS_2014}.
	
	The simplest example of ergodicity breaking in a many-body system is via the presence of integrability. It is assumed that a general model composed of several interacting particles is non-integrable, and possibly ergodic. However, within integrable models there exists several conserved quantities leading to the characterisation of each state with specific “good” quantum numbers \cite{Caux_Integrability_2011}. Another paradigm of ergodicity breaking is that of Many-Body Localisation (MBL), which generally appears in the presence of disorder and leads to emergent conserved quantities which induce localisation behaviour \cite{AbaninRMP}. The stability of MBL, particularly in a quantum chaotic model at the thermodynamic limit, has been previously discussed in literature \cite{QC_MBL, QC_MBL2} and underlines the difficulty in differentiating true ergodicity breaking from long-lived prethermal phases \cite{thermal_vs_prethermal, AbaninPrethermal, Ponte_Prethermal_2015}.
	
	The above mentioned features of integrability and MBL are examples of ergodicity breaking which affect all eigenstates of a given system, thereby implying that deviation from thermal behaviour is observed irrespective of the initial state undergoing unitary dynamics. Contrary to this, there also exist ergodicity breaking mechanisms which only affect specific eigenstates of the overall spectrum \cite{Moudgalya_2022}. This implies that ergodicity breaking, while present, is weaker as it affects only specific initial states undergoing unitary evolution. There exist several examples of the latter such as quantum scar states \cite{Turner_Scars_2018} and Hilbert space fragmentation \cite{Sala_Fragmentation_2020, Khemani_Fragmentation_2020}.
	
	A fundamental probe that is often used to differentiate between thermalising and ergodically broken models are spectral properties such as the level spacing statistics of the underlying Hamiltonian. Strong ergodicity breaking mechanisms such as integrability and MBL affect the entire energy spectrum leading to Poisson statistics, contrary to the Wigner–Dyson statistics observed for interacting and thermalising systems \cite{Bohigas_BGS, Mehta_RMT}. For the other class of weak ergodicity breaking, the overall Wigner–Dyson signature remains intact as only a very few states violate thermal behaviour, and therefore do not affect the entire level spacing distribution which originates from a significantly larger fraction of eigenstates. A change in spectral features also marks the breakdown of integrability upon tuning system parameters, such as in the transverse field Ising model where the introduction of a longitudinal magnetic field drives the system from integrable to chaotic dynamics \cite{Kim_Huse_TFIM_2013}. Further, there also exists additional evidence of shifts in spectral and dynamical properties on varying parameters such as the interaction strengths \cite{Levan_PRB} and asymmetric hopping terms \cite{Kartikeya_PRB}. 
	
	In our present work, we put forward results which also indicate a variation towards ergodicity breaking within the non-integrable and chaotic transverse field Ising model by adding inhomogeneity in the interaction strengths of spin–spin couplings within the chain. In order to explain our motivation for investigating this setup we bring it to the readers attention that changing the coupling parameter $J$ in the term $J\sum_iS^z_{i}S^z_{i+1}$ enhances the range of energy eigenvalues by a similar factor. This change is trivial in origin because the said term is diagonal in the computational basis and therefore directly affects the spread of the eigenvalue spectrum. Also, the level spacing statistics remain largely unchanged given the requirement of prior spectral unfolding to obtain such statistics \cite{Haake_QuantumChaos}. However, the level spacing can indeed be affected in a non-trivial manner by changing the coupling parameter J for only a part of the system. In particular, such a change could affect the energy spectrum in a system showing Wigner–Dyson statistics, leading to an uncorrelated spread of energy eigenvalues. In our present work, we show this to be indeed possible, and we present results indicating that the change in spectral features is accompanied by associated changes in dynamical behaviour as well, introducing suppressed ergodicity in what is otherwise a standard example of non-integrable chaotic dynamics.
	
	With our model, we observe a setup where the ergodic nature of the system is tunable upon changing the extent of inhomogeneity. We investigate the effects of this tunability on the dynamical behaviour of the system by studying operator spreading in both the real-space and Krylov bases \cite{Parker_Krylov_2019, Rabinovici_Krylov_2021}, while also analysing the growth of entanglement within the system \cite{Calabrese_Cardy_2005}. These diagnostics allow us to probe the nature and extent of ergodicity constrains introduced in the model, and comment on the usage of operator spreading as a reliable quantifier of the different dynamical classes.
	
	The manuscript is organized as follows: We shall introduce our setup of the inhomogeneous model in Sec. \ref{Model} while commenting on the mechanism responsible for the restrained ergodic nature and start with presenting a contrast between the early and late time dynamics, making use of the OTOC, as a primary quantifier in Sec. \ref{OTOC_Section}. We shall then present the variation in level spacing statistics in Sec. \ref{spectrum} before moving on the Krylov space description in Sec. \ref{krylov} and considering the entanglement of eigenstates in Sec. \ref{entanglement}. We conclude with our findings in Sec. \ref{conclusions}.
	
	\section{Inhomogeneous transverse field ising model}
	\label{Model}
	
	\begin{figure}[h]
		\centering
		\includegraphics[width = 0.5\textwidth]{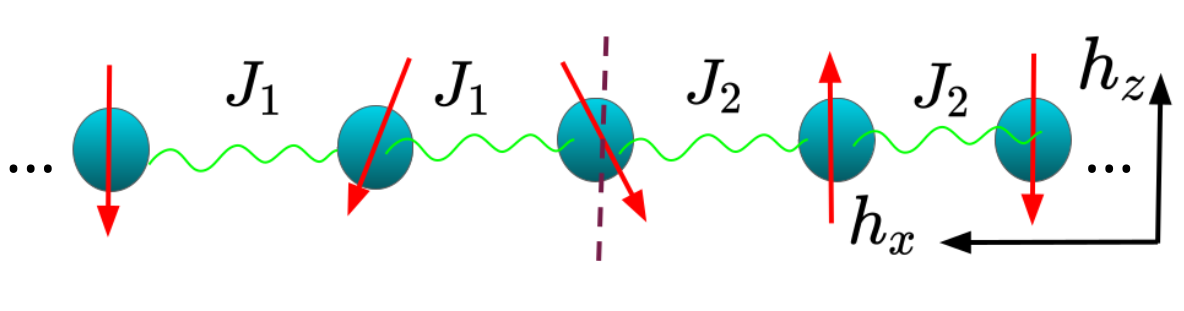}
		\caption{Schematic of an inhomogeneous spin chain with an odd number of spins, chosen so that the number of nearest-neighbor interaction terms can be equally divided. The spins interact along the $z$ direction with open boundary conditions. Half of the interaction terms have strength $J_1$ and the other half have strength $J_2$, creating a nonuniform interaction pattern. The chain is subjected to a transverse field $h_x$ and a longitudinal field $h_z$.}

		\label{Figure1}
	\end{figure}
	
	\begin{figure*}
		\includegraphics[width = 0.9\linewidth]{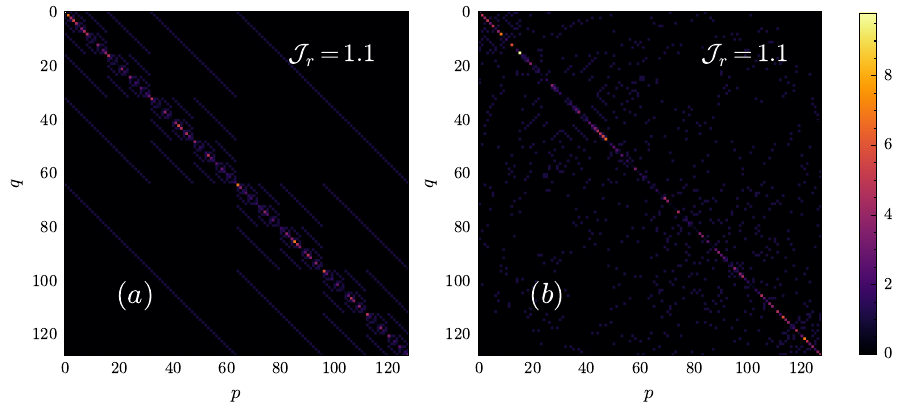}
		\includegraphics[width = 0.9\linewidth]{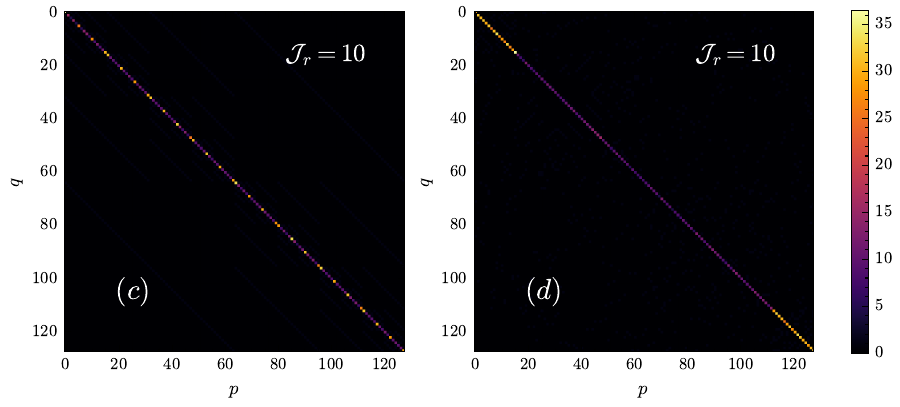}
		\caption{3D heat map of matrix elements, $H_{pq}$ for the Hamiltonian under different values of inhomogeneity parameter $\mathcal{J}_r$, for $N = 7$, $h_x = 1.05$ and $h_z = 0.5$. $(a)$ and $(c)$ are representations in the general computational basis, whereas $(b)$ and $(d)$ follow the basis of the dominant Hamiltonian term $H_0$. Note the suppression in the off-diagonal elements between $\mathcal{J}_r = 1.1$ and $10$.}
		\label{Figure2}
	\end{figure*}
	
	Central to our observations in the present work is the transverse field Ising model with an introduced notion of inhomogeneity. Here, we consider a system of $N$ spins, wherein the exchange interaction strength is given as $J_1$ in the first half of the system and $J_2$ for the latter half. This inhomogeneity is quantified by the parameter $\mathcal{J}_r$, given as the ratio $J_2/J_1$. Further, we consider the non-integrable version of the underlying transverse field Ising model, implying the presence of both longitudinal and transverse magnetic fields. Also, $N$ is taken as odd, so that the resulting ($N - 1$) bonds maybe equally divided between having values $J_1$ and $J_2$. The Hamiltonian is specified as:
	\begin{eqnarray}
		\hat{H} = -J_1 \sum_{i = 1}^{(N-1)/2} \hat{\sigma}^z_{i}\hat{\sigma}^z_{i+1} -J_2\sum_{i = (N+1)/2}^{N - 1} \hat{\sigma}^z_{i}\hat{\sigma}^z_{i+1} \nonumber \\
		-h_x\sum_{i=1}^{N }\hat{\sigma}^x_{i} - h_z\sum_{i=1}^{N}\hat{\sigma}^z_{i}. \,\,\,\,\,\,\,\,\,\,\,\,\,\,\,\,\,\,\,\,\,\,\,\,\,\
		\label{hamiltonian}
	\end{eqnarray}
	Here, we consider open boundary conditions which breaks the translational symmetry. Further, adding the inhomogeneity factor $\mathcal{J}_r$ breaks the parity symmetry which is defined as a reflection with respect to the central spin within the chain. This model, from the onset allows us to see the interplay of several factors such as symmetry breaking and nonintegrablility along with constrained transport properties within the system introduced by inhomogeneity. Also, an effect on quantum phase transitions is also observed \cite{abhijit_chaudhari_inhomogeneous}. The associated unitary is given as $\hat{U}(t) = \text{exp}(-i\hat{H}t)$ with $\hbar = 1$. 
	
	We find that changing the parameter $\mathcal{J}_r$ away from unity introduces effects similar to ergodicity breaking, and constraints the dynamical nature of the model. In order to find the physical mechanism underlining this effect we first observe that a large value of the parameter $\mathcal{J}_r$, implies that the term $J_2\sum_{i = (N+1)/2}^{N} \hat{\sigma}^z_{i}\hat{\sigma}^z_{i+1}$ takes on a dominant role in the overall Hamiltonian. This is because $J_2 = J_1\mathcal{J}_r$, making it the largest contributor to the Hamiltonian when inhomogeneity is higher. As this term is diagonal and generates an extensive set of local conserved quantities, we can claim that enhancing the value of $\mathcal{J}_r$ leads to an effectively integrable model \cite{Abhishek_Dhar_Gibbs, weak_integrability_breaking,Prethermal_effective_integrability, eigenstate_effective_integrability, prethermal_lucas}. This is further validated when we observe the effect on other dynamical indicators, which shall be presented in the following sections.
	
	Thus, it appears that increasing $\mathcal{J}_r$ produces a strong separation of energy scales, leading to approximate conservation laws and effective integrability \cite{BravyiSW,AbaninPrethermal,MoriPrethermal,KuwaharaPrethermal}.
	In this situation, although the full Hamiltonian does not admit an exactly solvable form, the dynamics is constrained by emergent quasi-conserved quantities or strong energy-scale separations. This leads to suppressed thermalisation, Poisson-like level statistics, and long-lived prethermal behavior \cite{DAlessioReview,AbaninRMP}.
	
	A natural framework to analyze effective integrability in the presence of a large separation of energy scales is provided by the Schrieffer--Wolff (SW) transformation \cite{SchriefferWolff,BravyiSW}. Writing the Hamiltonian as $H = H_0 + V$, with the dominant integrable term $H_0$ setting a large energy scale and a weaker perturbation $V$, the SW transformation systematically constructs an effective Hamiltonian which is approximately diagonal in the $H_0$ eigenbasis \cite{BravyiSW,MacDonaldSW}.
	Within this framework, perturbative off-diagonal couplings are algebraically suppressed in terms of $\lambda$, defined as the ratio between the perturbation strength and the dominant energy scale, consistent with controlled perturbative expansions in interacting quantum systems
	\cite{KehreinBook}. It is straightforward to identify the signature of a Schrieffer--Wolff transformation using a computational approach by analyzing the magnitude of off-diagonal elements of the Hamiltonian $H$ when written in the $H_0$ eigenbasis \cite{Morningstar}. Given the structure of the SW transformation, the off-diagonal contribution is expected to follow a scaling behavior of the form
	\begin{equation}
		|H_{\mathrm{off}}| \propto \lambda^{-\alpha}.
	\end{equation}
	In ideal perturbative limits, integer values of $\alpha$ are expected, corresponding to the order of the leading non-vanishing perturbation terms. However, in interacting many-body systems, effective non-integer exponents generically arise due to resonances, operator dressing, and mixed-order processes \cite{AbaninPrethermal,DeRoeckHuveneers}.
	
	To validate the suppression of off-diagonal elements in an intuitive manner, we evaluate the matrix elements of the Hamiltonian for different values of the parameter $\mathcal{J}_r$. Plotting these elements on a three-dimensional graph clearly reveals the dominant contribution of diagonal elements in the large-$\mathcal{J}_r$ regime (see Fig.~\ref{Figure2}), consistent with strong energy-scale separation
	\cite{SerbynLIOM}.
	This behavior follows from the term
	\begin{equation}
		H_0 = J_2 \sum_{i = (N+1)/2}^{N} \hat{\sigma}^z_{i}\hat{\sigma}^z_{i+1}, \,\,\,\,\, J_2 = \mathcal{J}_r \cdot J_1
	\end{equation}
	which constitutes the principal component $H_0$ and is diagonal in the computational basis.
	Representing the Hamiltonian in the $H_0$ eigenbasis yields the same qualitative observation. While this basis coincides with the computational basis, the eigenstates are ordered according to their absolute energy eigenvalues. In Fig.~\ref{Figure2}, we plot the Hamiltonian matrix elements in both bases for smaller and larger values of the inhomogeneity parameter $\mathcal{J}_r$. Having represented the Hamiltonian in this basis, we analyze the off-diagonal contribution as a function of $\mathcal{J}_r$. Although $\mathcal{J}_r$ does not correspond to an exact ratio between the principal and perturbative components due to the presence of transverse and longitudinal field terms, this parametrization nevertheless provides meaningful insight into effective integrability and the associated signatures of constrained ergodicity
	\cite{DAlessioReview}.
	\begin{figure}
		\centering
		\includegraphics[width=0.9\linewidth]{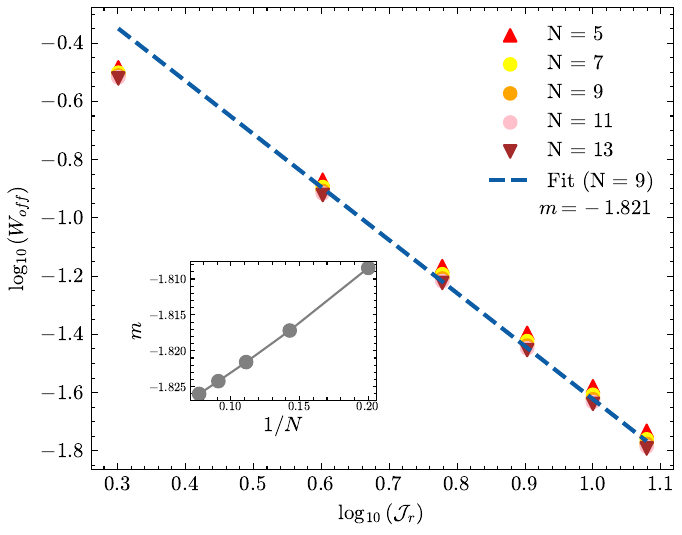}
		\caption{Logarithmic variation of normalised off-diagonal Hamiltonian components in the $H_0$ basis, $\sum_{i\neq j}|H_{ij}|$, with the homogeneity parameter for different values of $N$. Fitting a straight line, of slope $m$ between the plot of the logarithmic quantities indicates that $\sum_{i\neq j}|H_{ij}| \propto \mathcal{J}_r^{-1.821}$ for $N = 9$. The  inset shows variation of the fitted slope $m$ for different values of $1/N$. This curve indicates the extrapolated value of slope to be $-1.83$ for the large $N$ limit.}
		\label{Figure3}
	\end{figure}
	\begin{figure*}
		\includegraphics[width=0.95\textwidth]{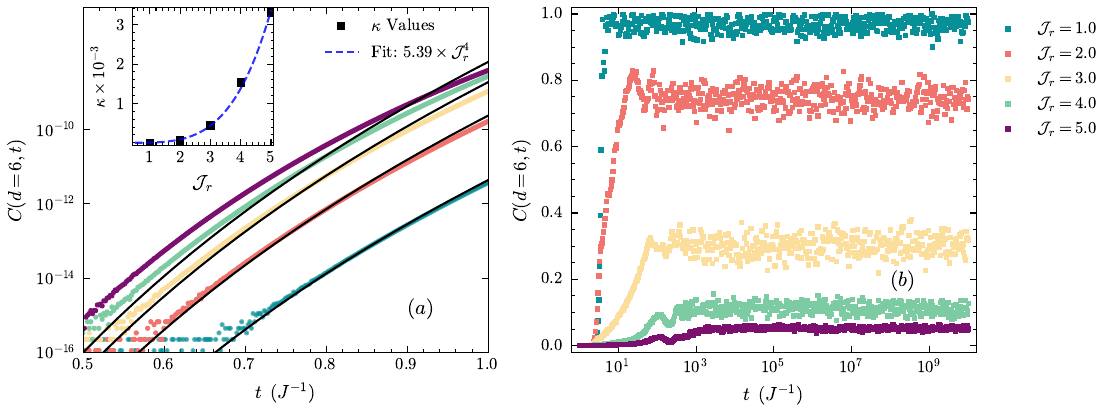}
		\caption{$(a)$ Variation of the OTOC (Eq. \ref{OTOC_Working}) in the short time limit for different inhomogeneity parameter $\mathcal{J}_r$. $h_x = 1.05$, $h_z = 0.5$ for $J_1 = 1$, $N = 7$ and $d = 6$ with open boundary conditions. The dotted lines indicate the fitted curve ($\propto t^{26}/13!$) for the corresponding $\mathcal{J}_r$. In $(b)$ we show the variation of parameter $\alpha$ which is found to be dependent on $\mathcal{J}_r$ as $5.39 \times \mathcal{J}_r^4$.}
		\label{Figure4}
	\end{figure*}	
	
	In order to quantify the contribution of off-diagonal components, we normalize the Hamiltonian matrix using the Frobenius norm, defining the normalized operator $\mathcal{H}$, which related to the Hamiltonian $H$ as:
	\begin{equation}
		\mathcal{H} = \frac{H}{\|H\|_F},
	\end{equation}
	where $\|H\|_F^2 = \sum_{i,j} |H_{ij}|^2$.
	This normalization ensures $\sum_{i,j} |\mathcal{H}_{ij}|^2 = 1$, allowing for a basis-independent and system-size--independent comparison across different parameter regimes \cite{LuitzChaos,MondainiChaos}. The off-diagonal weight is then defined as
	\begin{equation}
		W_{\mathrm{off}} = \sum_{\substack{i,j \\ i \neq j}} |\mathcal{H}_{ij}|^2.
	\end{equation}
	
	Once the Hamiltonian matrix is constructed and transformed into the $H_0$ eigenbasis, we plot $W_{\mathrm{off}}$ as a function of $\mathcal{J}_r$ on logarithmic scales (see Fig.~\ref{Figure3}). The resulting linear behavior, where the negative slope indicates a power-law suppression given as:
	\begin{equation}
		W_{\mathrm{off}} \propto \mathcal{J}_r^{-\alpha},
	\end{equation}
	with the exponent $\alpha \approx 1.8$.
	The magnitude of $\alpha$ increases with system size $N$, extrapolating to $\alpha \approx 1.83$ in the thermodynamic limit.
	This systematic suppression of off-diagonal couplings with increasing inhomogeneity is consistent with a near Schrieffer--Wolff regime and emergent effective integrability \cite{AbaninRMP,MoriPrethermal}. The deviation from the ideal perturbative value $\alpha = 2$ can be attributed to the presence of transverse and longitudinal magnetic field terms, which generate resonant and mixed-order processes beyond leading-order SW theory
	\cite{DeRoeckHuveneers}. Having established clear evidence of suppressed off-diagonal couplings and slow thermalization, we now turn to the computation of out-of-time-order correlators (OTOCs) to probe the dynamical consequences of effective integrability and constrained operator spreading in our model
	\cite{LarkinOvchinnikov,Swingle2018, SwingleTutorial}.

	\section{Initial versus late time dynamics} 
	\label{OTOC_Section}
	In order to evalute the dynamical properties of the system, we consider the OTOC, which serves as a quantitative probe of quantum information scrambling \cite{PRL_OTOC}. The metric captures how initially local operators spread and become highly nonlocal under unitary dynamics, thereby diagnosing scrambling and the growth of operator complexity. The Heisenberg evolution for an operator $O$ is given as: $O(t) = e^{i H t} \, O \, e^{-i H t}$. Following this, for two local spin operators $\sigma_i^\alpha$ and $\sigma_j^\beta$, acting on sites $i$ and $j$ respectively with $\alpha,\beta \in \{x,y,z\}$, the out-of-time-ordered correlator (OTOC) is defined as \cite{Fortes_OTOC}:
	\begin{equation}
		C_{i,j}^{\alpha,\beta}(t) 
		= - \big\langle \big[ \sigma_i^\alpha(t), \, \sigma_j^\beta(0) \big]^2 \big\rangle,
		\label{OTOC_General}
	\end{equation}
	where $\langle \cdot \rangle$ denotes the expectation value in a chosen ensemble. For the case of thermal state, the expectation value is reduced to the trace of commutator, which can then be written in terms of the four-point correlator \cite{Rohit_PRB}. Doing so, with the example of $\sigma_z$ observables at the sites $i$ and $j$ we get the following form:
	\begin{equation}
		C(d,t) =1- \text{Re}\Big[\tr\big( \hat \sigma_j^z(t) \hat \sigma_i^z(0)\hat \sigma_j^z(t) \hat \sigma_i^z(0)\big)\Big],
		\label{OTOC_Working}
	\end{equation}
	here, $d = |i - j|$, while the indices $\alpha$ and $\beta$ are dropped as both of them take the same value of $z$. Following the continuous time dynamics generated by the Hamiltonian given in equation Eq. \ref{hamiltonian} and writing the time evolved structure of operator $\sigma_j^z(t)$ we can find the power law time dependence of $C(d,t)$. In the non-integrable form (i.e. $h_x, h_z \neq 0$) the expression $C(d,t) \propto t^{2(2d+1)}/(2d + 1)!$, with the system parameters determining the exact proportionality constant \cite{Fortes_OTOC}. Given the structure of inhomogeneity, it becomes apparent that the effect of $\mathcal{J}_r$ only appears for $d \ge (N-1)/2$. In our results we consider $N = 7$, with the operators at extreme ends of the chain implying $d = 6$. We take the parameters $h_x = 1.05$ and $h_z = 0.5$ which gives a perfect Wigner-Dyson type distribution for the case $\mathcal{J}_r = 1$ and therefore highly ergodic dynamics.
	
	From a physical perspective, the system parameters play an explicit role in the early time regime, and this is reflected in the short-time growth of the OTOC \cite{Rohit_PRA}. Here, it is evident that the factor $\mathcal{J}_r^4$ appears in the expression of $C(d,t)$, which follows a variation involving the proportionality constant $\kappa$ as $\kappa(\mathcal{J}_r^4) \cdot t^{26}/(13)!$. This follows from the general expression mentioned above for $d = 6$. The functional form of $\kappa$ is approximately given by $5.39\mathcal{J}_r^4$, with the constant term being dependent on the numerical value of $h_x$ and $h_z$ (see Figure \ref{Figure4} $(a)$). The information presented in this Figure is obtained by plotting the variation of OTOC, given by Eq. \ref{OTOC_Working} for $t \in [0,1]$ with increasing values of $\mathcal{J}_r$. Then we fit the curve $t^{26}/13!$ to find the proportionality constant $\kappa$ corresponding to each value of $\mathcal{J}_r$. We then fit the obtained values of $\kappa$ corresponding to the curve $b\mathcal{J}_r^4$, revealing the constant $b$ to be $5.39$.
	
	The time variation for the OTOC, given a quantum chaotic spin Hamiltonian follows a clear trend of rapid power law growth during the initial phase (i.e. until Ehrenfest time) followed by a saturation behaviour around a numerical value determined by the equivalent ensemble of random matrices \cite{Rohit_PRB}. For the transverse field Ising model at $\mathcal{J}_r = 1$, it is known that $C(t) \to 1$ for large values of $t$, assuming that the parameters $h_x$ and $h_z$ lead to level separation described by the Wigner-Dyson distribution. This can be explained from the working formula given by Eq. \ref{OTOC_Working}, where $\sigma_j^z(t)$ takes the form of a random matrix for $t$ in the long-time limit. This leads to term subtracted from unity to gradually vanish as the convergence to random matrices is established, and thereby leading to the long-time value of the OTOC being unity.
	
	In our results we observe a significant deviation from the default variation of $C(t)$ when $\mathcal{J}_r$ exceeds the value of unity, as a result of the emerging slow thermalising behaviour. For parameters $h_x = 1.05$ and $h_z = 0.5$ it might be expected that the value of $C(t)$ approaches $1$ for longer times. However, we observe that for a higher value of $\mathcal{J}_r$ the OTOC saturates at increasingly lower values in the long time limit (see Figure \ref{Figure4} $(b)$). Interestingly, this behaviour is contrary to what is expected from symmetry arguments alone, where the growth rate of the OTOC should be higher with $\mathcal{J}_r \neq 1$ due to symmetry breaking, while saturating at a similar level. Taking into account the expression of $C(t)$ in the short time regime where it scaled with $\mathcal{J}_r$ as power law function, one aspect of the symmetry argument it true, however the long-time saturation behaviour is exactly opposite to similar apriori expectations.  
	
	We see that higher values of $\mathcal{J}_r$ show an enhanced growth rate for the OTOC in short-time regime, but saturate at a lower value in the long-time limit. This diminished saturation value is sustained for a large time period of the order $10^{10}$, indicating that the system remains in the pre-thermal phase within an extremely long time scale. This is however, not indicative of conventional integrable dynamics, where a persistent oscillation is observed \cite{Fortes_OTOC}, and the OTOC does not saturate. Further, with the magnetic field terms implying a nonintegrable model we can claim that inhomogeneity introduced by taking $\mathcal{J}_r \neq 1$ induces suppressed ergodicity in the dynamics associated with the transverse field Ising model, as expected in the regime of effective integrability. This correlates well with the contrasting observations for fast and slow scramblers \cite{Swingle2018,Nahum2018} even beyond clean systems \cite{Fan2017,He2017,Swingle2018} in the regime of many-body localisation. Further, OTOCs have also been obtained experimentally in a wide array of platforms, including trapped ions~\cite{Garttner2017}, NMR systems~\cite{Li2017}, superconducting qubits~\cite{Mi2021, Nature_OTOC_2025}, and trapped ions~\cite{Blatt2012}. With this we can observe the relevance of OTOC as a robust and experimentally viable metric, which also gives a clear quantification for the different dynamical properties exhibited by our model.  
	
	\section{Variation of Spectral Properties}
	\label{spectrum}
	\begin{figure}
		\includegraphics[width=\linewidth]{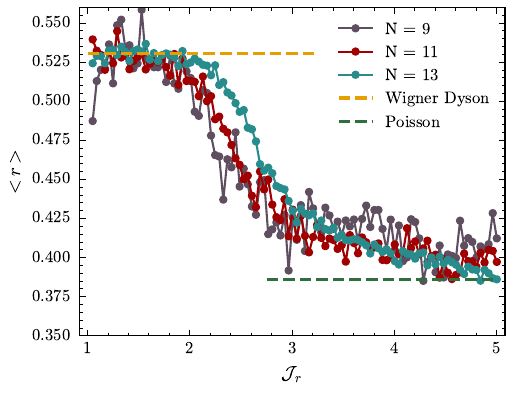}
		\caption{Variation of the average spectral parameter $r$ for different values of $\mathcal{J}_r$, with system size appearing as a parameter. For all cases, $h_x = 1.05$ and $h_z = 0.5$ with open boundary conditions. The level spacing includes the entire spectrum as a particular symmetry segment does not exist.}
		\label{Figure5}
	\end{figure}
	
	In our analysis so far we have observed signatures of suppressed ergodicity and slow thermalisation in the inhomogeneous model, based on the saturation behaviour of the OTOC. In order to validate our analysis we shall now present results regarding the spectral properties of the model and its dependence on $\mathcal{J}_r$. It is well known that quantum chaotic models show level spacing described by the Wigner-Dyson distribution, while for an model showing features of effective integrablility, the same follows Poissionian statistics \cite{Bohigas_BGS, AbaninPrethermal}.
	
	In order to study the level spacing statistics, we consider the value of parameter $\mathcal{J}_r$ between $1.05$ and $5.0$. Since for all values of $\mathcal{J}_r \neq 1$, the parity symmetry is already broken and hence, the step of block diagonalising the Hamiltonian in a particular symmetry segment is not required. Once the eigenvalues are obtained and sorted we carry out the process of 'unfolding' which ensures the mean level spacing is equal to unity. Once this is done, the distribution is plotted and found to transition from clearly a Wigner-Dyson to Poission distribution. To show this variation, we consider the $r$ parameter as a spectral measure, the numerical value of which is indicative of chaos. The parameter is defined as follows \cite{KC_Chaos}, given $s_i = E_i - E_{i-1} (> 0)$ is the energy level spacing between the $i$th and $(i - 1)$th eigenvalues:
	\begin{equation}
		r = \frac{1}{2^N} \sum_{i=1}^{2^N - 1} \tilde{r}_i; \,\,\,\,\ \text{where} \,\,\,\,\ \tilde{r}_i = \frac{\text{min}(s_i,s_{i-1})}{\text{max}(s_i,s_{i-1})}.
	\end{equation}    
	When defined in this manner, the value of $r \approx 0.535$ for the Wigner-Dyson distribution and $r \approx 0.386$ for Poissionian. Note that these exact values are obtained for a large system size, however the values remains approximately true \cite{Levan_PRB}. In our result (see Figure \ref{Figure5}) we plot the obtained values of $\langle r \rangle$ for different values of $\mathcal{J}_r$ at $50$ equally spaced points in the interval $[1.05,5]$ and a clear transition can be observed. The data set used here corresponds to $N = 9,11$ and $13$, with the larger system size leading to a more uniform variation and with suppressed fluctuations. Note that $\mathcal{J}_r = 1$ is not plotted in this graph as the process of evaluating that must include the aforementioned step of block diagonalising to a specific symmetry segment. Moreover, for our parameters it is known that $r = 0.53$ for $\mathcal{J}_r = 1$ \cite{BGS2}. 
	
	While the $r$ parameter is static in nature, a dynamical metric which captures the notion of ergodicity breaking via spectral properties is given as \cite{QC_MBL, QC_MBL2}:
	\begin{equation}
		g = \text{log}_{10}\bigg( \frac{t_{H}}{t_{Th}} \bigg),
	\end{equation} 
	where $t_H$ and $t_{Th}$ are respectively the Heisenberg and Thouless time. The metric $g$ is logarithm of dimensionless conductance $t_{Th}/t_{H}$ which is used to analyse transport properties in an Anderson localised system. An advantage of this metric, apart from it being numerically easy to obtain, is that a clear transition from ergodic to localised dynamics can be found when $t_{Th} \approx t_{H}$. $g$ is obtained as a scaling solution of system size and disorder, thereby being indicative of the universal transition behaviour. Due to this, the parameter $g$ serves as a useful metric for classifying ergodicity breaking in a finite system, extending into the Many-Body localised (MBL) regime \cite{QC_MBL}. 
	
	In our model we do not have an explicit notion of disorder in the system, but a clear trend is observed towards slow thermalisation upon varying the parameter $\mathcal{J}_r$. For this reason we study the variation of $g$ presenting an alternative analysis where inhomogeneity takes on the role of suppressing ergodicity instead of disorder strength. While the parameter $g$ is not typically defined for such a disorder free framework given its relation to conductance, it nonetheless gives us an idea of the extent of ergodicity breaking within the system dynamics allowing us to compare the same against an example where the said effects are driven by disorder \cite{QC_MBL}. 
	
	A greater significance towards quantifying the dynamics in terms of ergodicity breaking is via the Thouless time $t_{Th}$, marking the onset of ergodic behaviour within the system \cite{DAlessioReview}.  It can be extracted from the Spectral Form Factor (SFF), being the point where it begins to overlap with the predicted values of the corresponding random matrix ensemble. We thus evaluate the SFF for our system with increasing value of $\mathcal{J}_r$, following the procedure outlined in \cite{QC_MBL}. The SFF is defined as:
	\begin{equation}
		SFF(\tilde{t}) = \frac{1}{Z} \left\langle |\sum_{\alpha = 1}^{2^N} \rho(\epsilon_{\alpha}) e^{-i2\pi\epsilon_{\alpha}\tilde{t}}|^2 \right\rangle_{t_w},
		\label{SFF_Eq}
	\end{equation} 
	where $\tilde{t}$ denotes the scaled time and $Z$ is a normalisation factor such that $SFF(t >> 1) \approx 1$. The average $\langle \cdot \rangle_{t_w}$ is carried out over a moving time window and becomes important in constraining the rapid fluctuations that arise in numerically evaluating the metric. 
	\begin{figure}
		\includegraphics[width=\linewidth]{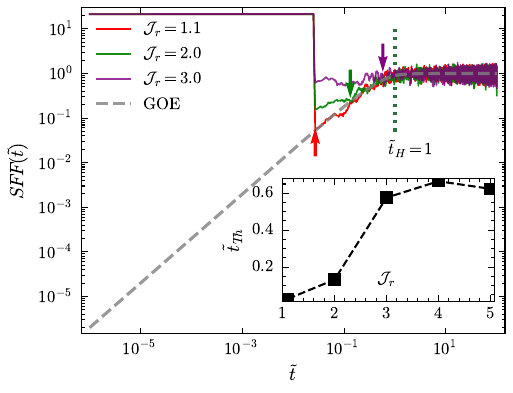}
		\caption{Variation of spectral form factor for different values of $\mathcal{J}_r = 1.1,2,3$, in scaled time units $\tilde{t}$ such that $\tilde{t}_H = 1$. The intersection with GOE is shown with markers. The scaling of $\tilde{t}_{Th}$ with $\mathcal{J}_r$ is given in the inset. (Data for $SFF(\tilde{t})$ at $\mathcal{J}_r = 4,5$ generated but not shown to avoid clustering of markers)} 
		\label{Figure6}
	\end{figure}
	
	An important consideration is that of the scaled time $\tilde{t}$, which is defined in a manner wherein Heisenberg time $t_H$ becomes unity. Heisenberg time is defined as the inverse of mean energy spacing, $t_H = \hbar/\delta\bar{E}$, where $\delta\bar{E}$ is the average over all $s_i$. With $\hbar = 1$ we achieve the scaled time taking the unfolded spectrum $\{\epsilon_1, \epsilon_2,...\epsilon_{2^N} \}$ such that for $\tilde{s}_i = \epsilon_i - \epsilon_{i - 1}$, and an average over all $\tilde{s}_i$ equals unity. The process is already carried out in our case while calculating the parameter $r$. For this, the steps required are as follows: Starting with the sorted list of energy eigenvalues $\{E_i\}$ we define the cumulative spectral function \cite{QC_MBL, unfolding_ashraf}:
	\begin{equation}
		G(\lambda - \lambda_i) = \sum_i \Theta(\lambda - \lambda_i)
	\end{equation}
	where $\Theta$ is the unit step function. We now fit a $n$ degree polynomial $g_n(\lambda)$, with $n = 10$ in our case. The unfolded spectrum $\{\epsilon_i\}$ is obtained as $g_n(E_i)$ where mean spacing $\delta\bar{\epsilon} = 1$ over all $\tilde{s}_i$. Thereby using $\{ \epsilon_i\}$ in Eq. \ref{SFF_Eq} implies the involvement of scaled time, where $\tilde{t}_H = 1$. 
	
	The final quantity mention in Eq. \ref{SFF_Eq} is $\rho(\epsilon_{\alpha})$ which is a gaussian filter to remove the effects arising out of spectral edges. It is defined as:
	\begin{equation}
		\rho(\epsilon_{\alpha}) = \text{exp}\left( -\frac{(\epsilon_{\alpha} - \bar{\epsilon})^2}{2\eta^2\Gamma^2} \right)
	\end{equation}
	where $\bar{\epsilon}$ and $\Gamma^2$ are the average and variance of $\{ \epsilon_i \}$ for the corresponding value of $\mathcal{J}_r$. The parameter $\eta$ specifies the number of effective states used, and is fixed to $0.5$. For the purpose of normalisation, we take $Z = |\sum_{\alpha} \rho(\epsilon_{\alpha})|^2$ to have $SFF(t >> 1) = 1$. Finally, we take a moving time average to stabilise the results. The value of SFF is calculated at $10^6$ equally spaced points between $10^{-6}$ and $100$, while the time window $t_w$ is taken as $51$. 
	
	Having the variation of $SFF(\tilde{t})$ we may now obtain $t_{Th}$ in the same units by obtaining the point of intersection for variation with the same for corresponding random matrix ensemble. Our system corresponds to the gaussian orthogonal ensemble ($\beta = 1$ via Dyson Classification) as the Hamiltonian is a real symmetric matrix. The SFF variation for this ensemble is given as:
	\begin{equation}
		SFF_{GOE}(\tilde{t}) = 2\tilde{t} - \tilde{t}\ln(1 + 2\tilde{t})
	\end{equation}  
	The overlap of $SFF(\tilde{t})$ with $SFF_{GOE}(\tilde{t})$ at $t_{Th}$ implies the onset of universal linear ramp, which is indicative of the onset of ergodic dynamics \cite{sff_review, sff_measure}.
	
	In our case progressively increasing the inhomogenity parameter $\mathcal{J}_r$ increases $t_{Th}$, indicating a delayed onset of ergodic dynamics resulting from weaker chaotic behaviour. For eg:, in Fig \ref{Figure6}, at $\mathcal{J}_r = 1.1$, we have $t_{Th} \approx 0.1$ which increases to $0.6$ for $\mathcal{J}_r = 5$ implying a 6-fold increase in $t_{Th}$ ($t_H$ remains fixed at $1$). For an analogous model described with disorder, this change corresponds to $g$ being transformed from a value around $10$ to $0.15$, which is near to an MBL transition, and associated with significant disorder strength. In any case, we observe significantly diminished chaotic behaviour in our model upon increasing $\mathcal{J}_r$. 
	
	With the insights using spectral diagnostics, we shall now observe the effect of this constrained ergodicity on the spread of operators, using Krylov complexity.
	\begin{figure*}
		\includegraphics[width=0.95\linewidth]{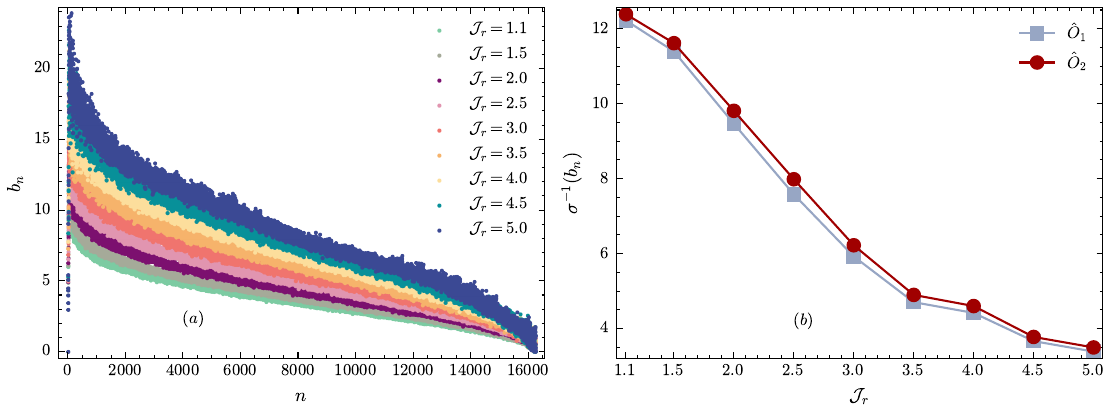}
		\caption{$(a)$ Variation of Arnoldi coefficients $\{ b_n\}$ over all iterations $n$, with the inhomogeneity factor $\mathcal{J}_r$ appearing as a parameter, for $\hat{O} = \hat{S}_z^1 + \hat{S}_z^7$ denoted as Operator 1 in the main text for system size $N = 7$. The same variation is obtained for Operator 2 as well and is not shown due to its similar behaviour. $(b)$ Variation of the standard deviation inverse $\sigma^{-1}(b_n)$ with parameter $\mathcal{J}_r$ for Operators 1 and 2. Note the increase in deviation with the breakdown of ergodic dynamics and the similar behaviour between both the initial operators.}
		\label{Figure7}
	\end{figure*}
	
	\section{Krylov Space Dynamics}
	\label{krylov}
	
	It is understood that under unitary dynamics the operator $\hat{O}$ expands its support and tends to explore a subspace of all possible operators, called as the Krylov space for the given system. For this subspace a suitable basis can be formed using the time-evolved structure of the operator and making use of the Lanczos algorithm. For the Hamiltonian given by $\hat{H}$, we have $\partial_t \hat{O} = i[\hat{H},\hat{O}]$; following the Heisenberg picture where \cite{HashimotoKrylov2023, Krylov_Standard_Paper, krylov_review}:
	\begin{equation}
		\hat{O}(t) = e^{i\hat{H}t} \hat{O} e^{-i\hat{H}t} = \sum_{n = 0}^{\infty} \frac{(it)^n}{n!} \mathcal{L}^n(\hat{O}),
	\end{equation}
	with $\mathcal{L}(\hat{O}) = [\hat{H},\hat{O}]$ being the Liouville operator. From the above expression it becomes apparent that,
	\begin{equation}
		\hat{O}(t) \in \text{Span} \{ \mathcal{L}^0(\hat{O}), \mathcal{L}^1(\hat{O}), ... , \mathcal{L}^{K-1}(\hat{O})\}.
	\end{equation} 
	This $K$ dimensional space is the aforementioned Krylov space, where $K$ in the case of considering the spread of operators is bounded above as \cite{Krylov_Standard_Paper}:
	\begin{equation}
		K \le D^2 - D + 1, 
	\end{equation}
	where $D$ represents the dimensionality of our system. The set of operators corresponding to the $D$ dimensional space form a Hilbert space of their own, with the role of inner product being taken by the Hilbert-Schmidt product:
	\begin{eqnarray}
		\langle A|B \rangle = \frac{1}{D}\tr[A^{\dagger}B]; \\
		|| A ||^2 = \langle A|A \rangle = \frac{1}{D} \tr[A^{\dagger}A].
	\end{eqnarray}
	With this definition of the inner product, we can carry out a procedure analogous to the Gram-Schmidt process in quantum mechanics to have an orthonormal basis for the set $\text{Span}\{ \mathcal{L}^0(\hat{O}),\mathcal{L}^1(\hat{O}),...,\mathcal{L}^{K-1}(\hat{O})\}$ given by $\{ \mathcal{W}_i\}_{i = 0}^{K - 1}$. The time evolved operator at time $t$ may now be expressed in terms of the Krylov basis vectors:
	\begin{equation}
		\hat{O}(t) = \sum_{n = 0}^{K - 1} i^n \beta_n(t) \mathcal{W}_n.
	\end{equation}
	The related metric of Krylov Complexity, which quantifies the spread of an operator in terms of the Krylov basis vectors is given as \cite{Krylov_Nandy, Krylov_Standard_Paper}:
	\begin{equation}
		\mathcal{K}_{C}(t) = \sum_{n = 0}^{K - 1} n|\beta_n(t)|^2.
	\end{equation}
	The metric of Krylov complexity also has an interpretation for the quantum circuit picture \cite{Nielsen_Krylov_PRL, Krylov_from_circuits, Krylov_unitary}, although it  has been shown to not be a distance measure \cite{Nielsen_Krylov_PRD}. In order to implement the Lanczos algorithm in the operator space we make use of Arnoldi iterations which involve the explicit orthogonalisation of each new vector with respect to all previously created vectors for numerical stability. The explicit steps followed are:
	\begin{itemize}
		\item $\ket{\mathcal{W}_0} = \ket{\hat{O}}/||\hat{O}||$. We also have $b_0 = ||\hat{O}|| = \langle \hat{O} | \hat{O} \rangle^{1/2}$.
		\item for $n \ge 1$, $\ket{\hat{A}_n} = \ket{\mathcal{L}(\hat{O}_{n-1})}$, where $\hat{A}_n$ is: \\
		$\ket{\hat{A}_n} = \ket{\mathcal{L}(\hat{O}_{n-1})} - \sum_{m = 0}^{n-1} \ket{\hat{O}_m}\langle{\hat{O}_m}|\hat{A}_n\rangle$.
		\item $b_n = \langle \hat{A}_n | \hat{A}_n \rangle^{1/2}$. These are the Arnoldi coefficients, which must be stored after each iteration.
		\item $\ket{\mathcal{W}_n} = \frac{1}{b_n}\ket{\hat{A}_n}$, if $b_n \neq 0$, else: Terminate. These are the Krylov basis vectors. These must also be stored at the end of each iteration.
	\end{itemize}
	Therefore at the end of all iterations we end up with the set of Krylov basis $\{ \mathcal{W}_i \}$ and the Arnoldi coefficients $\{ b_n \}$. Note that $\ket{\hat{X}}$ represents the operator $\hat{X}$ as a member of the Hilbert space that is formed of all operators and the corresponding inner product.

	Given the tridiagonal structure of the Liouville operator in the Krylov basis, the Arnoldi coefficients $\{ b_n \}$ map to the hopping amplitudes in a $1D$ tight-binding model. Thus the Krylov complexity $\mathcal{K}_C(t)$ can be considered as the spread of a given operator in the Krylov space, and this spread can be directly quantified in terms of the deviations within coefficients $\{b_n\}$. It can be understood as follows in relation to the ergodic nature of the dynamics: A thermalising evolution leads to a much greater spread of the operator, implying a regular distribution within the hopping coefficients $\{b_n\}$. In contrast, an integrable or otherwise ergodically broken model which shows a restrained operator spread must have irregularly spaced hopping coefficients, involving a wide range and irregular behaviour within the Arnoldi coefficients $\{b_n\}$ \cite{Krylov_Standard_Paper, Krylov_bala}.
	
	Such a spread within the set of coefficients can be evaluated by computing the deviations with respect to a moving local average. The exact computations performed are given by the following formula for a sequence of length $N$ and local width $w$ \cite{Wisniacki_average}:
	\begin{equation}
		\sigma^2(b_n) = \frac{1}{N}\sum_{n=n_0}^{N}(b_n - \bar{b}_n)^2, \text{where}, \bar{b}_n = \frac{1}{2w}\sum_{m = n-w}^{n+w} b_m
	\end{equation} 
	here $n_0$ represent the start of sequence and $N$ is given as $D^2 - D - w + 1$. For generating our results, we have taken $N = 7$, which implies a Hilbert space dimension of $D = 2^7 = 128$. Corresponding to this, the maximum dimension of the Krylov space is given by $16,257$. Given this number, we have taken $n_0 = 100$ and $w = 400$ for calculating the deviations described by the above expressions. Note that the high dimension of the Krylov basis appears because of our model lacking any specific symmetry segment arising out of translational or parity transformations. Given the structure of our model we take the normalised versions of two different operators viz. $\hat{O}_1 = \hat{S}_z^{1} + \hat{S}_z^{N}$ and $\hat{O}_2 = \hat{S}_z^{1} + \hat{S}_z^{(N-1)/2}$ so that we may explore their spread relative to each other in the Krylov space. The Arnoldi coefficients arising from the said operators and the said measure of deviation are presented in Figure Fig. \ref{Figure7}. 
	
\begin{figure}
		\includegraphics[width=0.96\linewidth]{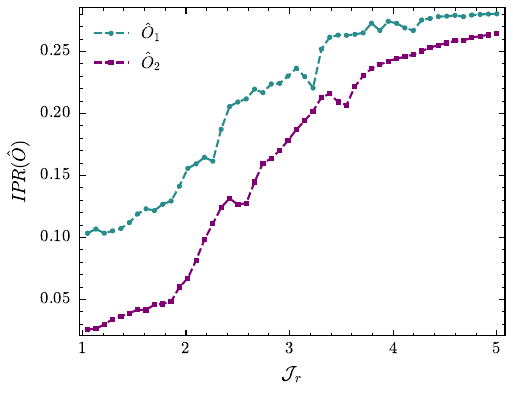}
		\caption{Variation of IPR for the initial operators $\hat{O}_1$ and $\hat{O}_2$ upon changing the inhomogeneity parameter $\mathcal{J}_r$.}
		\label{Figure8}
	\end{figure}

	The spread of coefficients $\sigma(b_n)$ along with the measure $\mathcal{K}_C(t)$ can be used as diagnostics to determine the nature of underlying dynamics. Within these the former has appeared to be a stronger and more reliable indicator than the latter, which is dependent on several factors other than dynamics such as initial operators and non-local nature of the same \cite{Krylov_multiseed, Krylov_sahu_nonlocal}. Moreover, the saturation value of Krylov complexity have been shown to be reversed for the same initial operators under different parameters \cite{KC_Chaos}. This behaviour is found to be driven by the initial spread of an operator in the Hamiltonian eigenbasis and is quantified by the inverse participation ratio (IPR) in the operator space. This has been shown to significantly influence the final saturation behaviour, to an extent that is greater than the dynamical nature of the system \cite{Aravinda_Sir_PRE}. In light of these observations one must be cognizant of the properties associated with the intial operators while exploring the growth and saturation of Krylov complexity for characterising the dynamical behaviour. In our present case we already have evidence of slow thermalisation on varying the inhomogeneity parameter $\mathcal{J}_r$ using other independent diagnostics. We can therefore explore the associated Krylov dynamics using our example exploring the departure from regular thermalising behaviour. We begin by specifying the IPR for our chosen initial operators upon changing the inhomogeneity factor $\mathcal{J}_r$ (see Figure Fig.\ref{Figure8}), before presenting the results for Krylov complexity $\mathcal{K}_C(t)$ with $O$ being taken as $\hat{O}_1$ and $\hat{O}_2$.
	\begin{figure*}
		\includegraphics[width=0.95\linewidth]{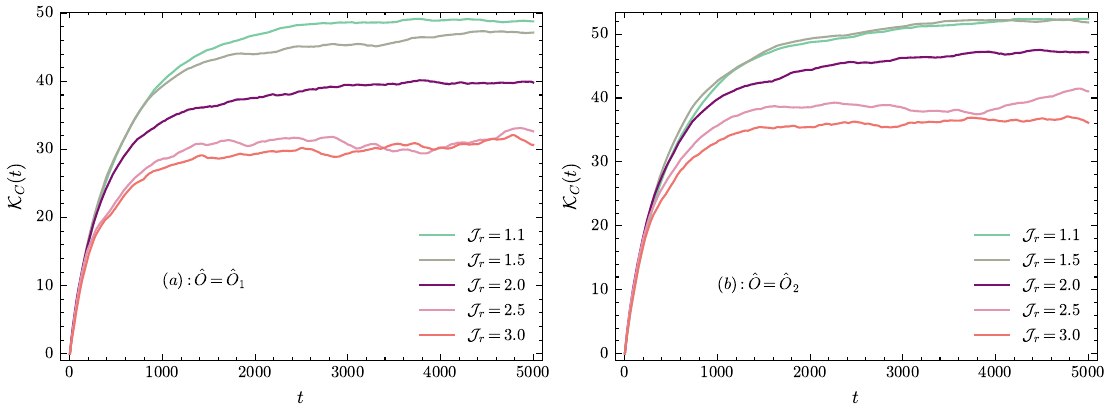}
		\caption{Variation of Krylov Complexity $\mathcal{K}_C(t)$ with time, for intial operators $(a)$ $\hat{O}_1$ and $(b)$ $\hat{O}_2$, for increasing values of the inhomogeneity parameter $\mathcal{J}_r$ in the thermalising regime. For the initial operators we consider, IPR increases monotonically with $\mathcal{J}_r$. It can be therefore observed that an increased IPR consequently leads to lower saturation value of $\mathcal{K}_C(t)$.}
		\label{Figure9}
	\end{figure*}  
	\begin{figure*}
		\includegraphics[width=0.95\linewidth]{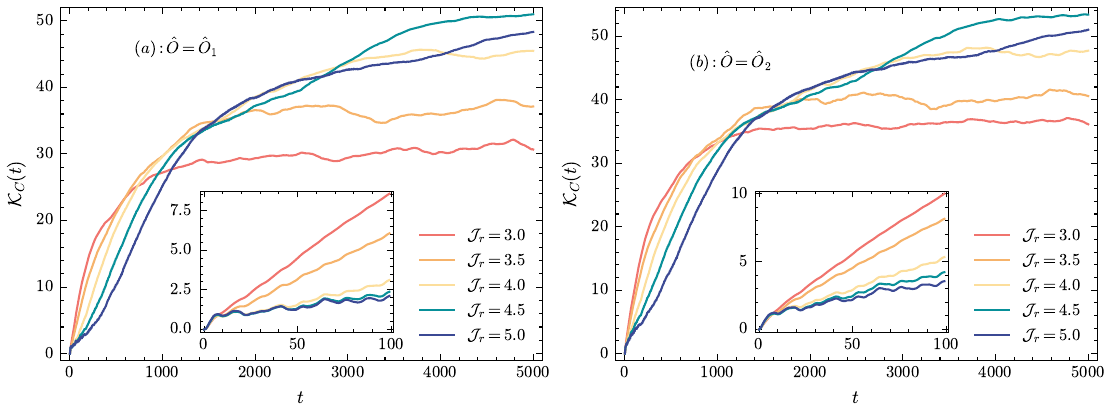}
		\caption{Variation of Krylov Complexity $\mathcal{K}_C(t)$ with time, for initial operators $(a)$ $\hat{O}_1$ and $(b)$ $\hat{O}_2$, for increasing values of the inhomogeneity parameter $\mathcal{J}_r$ in the regime with ergodicity breaking. Inset shows the variation zoomed into the short-time region. As before higher values of $\mathcal{J}_r$ lead to increased IPR, but this instead leads to a higher saturation value of $\mathcal{K}_C(t)$.}
		\label{Figure10}
	\end{figure*}  
	
	For a normalised operator $\hat{O} = \hat{O}/||\hat{O}||$, the spread in a Hamiltonian eigenbasis $\{ \phi_i \}$ can be quantified by the IPR, given as:
	\begin{equation}
		\text{IPR}(\hat{O}) = \sum_i |\langle \phi_i|\hat{O}|\phi_i\rangle|^2. 
		\label{IPR_Eq}
	\end{equation}
	In our case with the initial operators being $\hat{O}_1$ and $\hat{O}_2$, the IPR changes with a change in the value of $\mathcal{J}_r$. This is because for each value of $\mathcal{J}_r$ there is a distinct eigen spectrum with different eigenvectors. Therefore traversing through a set of $\mathcal{J}_r$ values induces a change in the IPR for operators $\hat{O}_1$ and $\hat{O}_2$. We find that for both cases the IPR increases monotonically on increasing $\mathcal{J}_r$. It has been explicitly shown for the Floquet class of models that a initial state with higher values of IPR give a lower saturation value of Krylov complexity $\mathcal{K}_C(t)$, irrespective of system dynamics. This observation indicates that initial operators play an important role in the measure of $\mathcal{K}_C(t)$, often superseding the role played by the dynamical nature of the system \cite{Aravinda_Sir_PRE}. Note that other examples of state dependent metrics used for detecting chaotic dynamics such as OTOCs and entanglement generation do not show a similar dependence on the initial state \cite{Aravinda_Sir_PRE}. 
	\begin{figure*}
		\includegraphics[width=0.95\linewidth]{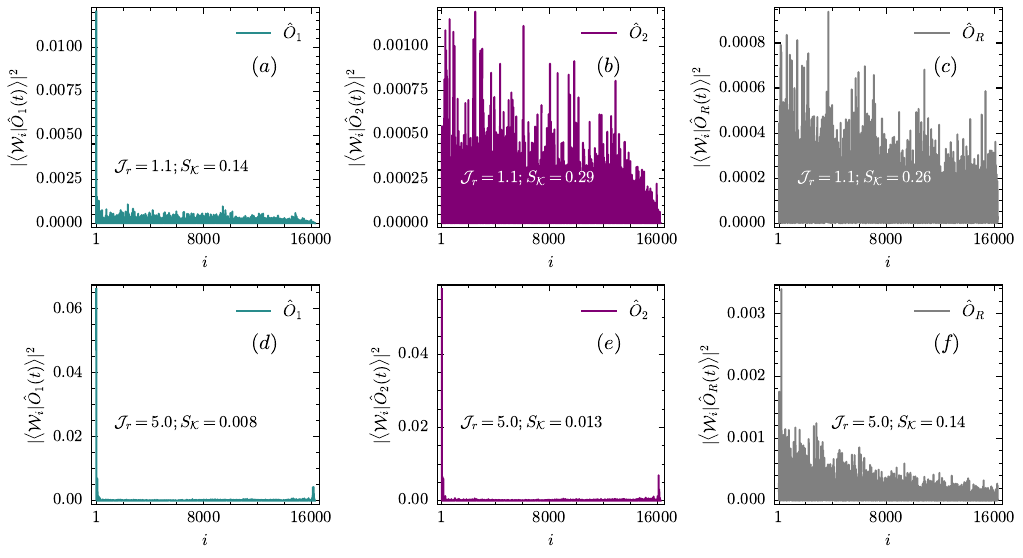}
		\caption{Overlap of the different time evolved operators at $t_{max} = 5000$ with the corresponding Krylov basis vectors for different ergodic regimes. For $(a-c)$ the value of $\mathcal{J}_r = 1.1$, while for $(d-f)$ the value of $\mathcal{J}_r = 5.0$. } 
		\label{Figure11}
	\end{figure*}
	
	In our results for Krylov complexity we obtain a similar IPR dependent behaviour in the regime where our system shows thermalising nature. However, when once we begin to obtain signatures of ergodicity breaking, the IPR dependence of $\mathcal{K}_C(t)$ is reversed. Nonetheless, it remains possible predict the saturation values of Krylov complexity based on properties of the initial operator (see Appendix \ref{IPR_Calculation}). In Figure Fig. \ref{Figure9} we calculate the measure $\mathcal{K}_C(t)$ taking $t$ to a maximum value of $5000$. We observe saturation behaviour at around $t = 1000$. Note that based on our previous results, we know that for $\mathcal{J}_r \in [1,3]$, the system shows thermal behaviour and in Fig. \ref{Figure9} the values of $\mathcal{J}_r$ for which the variation is shown belong to this particular regime. Considering $\hat{O}_1$, taking $\mathcal{J}_r$ from $1.1$ to $3$ increases the IPR from $0.1$ to $0.22$, and this suppresses the saturation value of $\mathcal{K}_C(t)$ from around $50$ to $25$ respectively. Similarly for $\hat{O}_2$ the IPR changes from $0.03$ to $0.17$ for the same change in $\mathcal{J}_r$, leading to the same suppression in $\mathcal{K}_C(t)$. For both cases, the initial growth of $\mathcal{K}_C(t)$ remains the same, irrespective of $\mathcal{J}_r$. The overall picture changes when considering the variations observed in Figure Fig. \ref{Figure10}. Here, we again plot $\mathcal{K}_C(t)$ varying $t$ till $t = 5000$. Firstly, it is clear that saturation behaviour takes a lot longer to be established. Most of the lines show a growing nature till $t \approx 4000$, with the variation for $\mathcal{J}_r = 5$ still growing at $t = 5000$. This is expected, and can be attributed to the gradual spread of the initial operators in the Krylov space, associated with the dynamics of slow thermalisation. Secondly, the IPR dependence of $\mathcal{K}_C(t)$ is reversed, with higher values of IPR giving greater $\mathcal{K}_C(t)$. For the case of $\hat{O}_1$ taking the value of $\mathcal{J}_r$ from $3$ to $5$ increases the IPR from $0.22$ to $0.3$ and this is accompanied by an increase in the saturation value of $\mathcal{K}_C(t)$ from $30$ to a value seemingly greater than $50$. Same is observed for $\hat{O}_2$, where IPR increases from $0.17$ to $0.25$, increasing the saturation value of $\mathcal{K}_C(t)$ from $35$ to $50$ upon increasing $\mathcal{J}_r$ from $3$ to $5$. However, the most important feature is the suppression of initial growth for $\mathcal{K}_C(t)$ when the ergodic nature of the dynamics are suppressed. Moreover, there is a systematic suppression of initial growth as the value of $\mathcal{J}_r$ increases and the effects of ergodicity breaking get stronger. This feature was absent when we were considering $\mathcal{J}_r$ to be in the ergodic regime, as growth was same, irrespective of $\mathcal{J}_r$. This allows us to comment that while the saturation value of $\mathcal{K}_C(t)$ has a high initial state dependence, the initial growth phase of the quantity captures system dynamics to some extent, particularly when involving some degree of ergodic suppression.
	
	To get a better understanding of the operator spread we generate the time-evolved version of our initial operators, and then find the overlap coefficients given by $\langle \mathcal{W}_i|\hat{O}(t = t_{sat})\rangle$ for all $i$. For our case, the total number of $\ket{\mathcal{W}_i}$ are $16,257$ and we take $t_{sat} = 5000$ for each case. The overlap values are shown in Figure Fig. \ref{Figure11}. The sum of all the values, or the total area of the shaded region in each of the graphs in Fig. \ref{Figure11} equals unity. To quantify the spread of time-evolved operators we introduce the measure $S_\mathcal{K}$, which is defined as:
	\begin{equation}
		S_{\mathcal{K}} = \frac{1}{K} \bigg[\sum_{i} |\langle \mathcal{W}_i|\hat{O}(t_{sat})\rangle|^4 \bigg]^{-1}
	\end{equation}
	This expression is exactly same as that used for operator space IPR, and to differentiate this measure from Eq. \ref{IPR_Eq} we use the different notation given by $S_{\mathcal{K}}$. $K$ is the maximum dimension of the Krylov space which is $16,257$. The quantity $\sum_{i} |\langle \mathcal{W}_i|\hat{O}(t_{sat})\rangle|^4$ is confined between $1$ and $K$ corresponding to the overlap of any one or all of the Krylov basis vectors $\ket{K_i}$ with $\hat{O}(t)$. This implies $S_{\mathcal{K}} \in [1/K,1]$. The lower limit is of the order $10^{-5}$ and maybe taken as zero. Thus, a high value of $S_{\mathcal{K}}$ indicates that a larger number of Krylov basis states have an overlap with the time evolved operator.
	
	In our results (see Fig. \ref{Figure11}) we find that for both operators $\hat{O}_1$ and $\hat{O}_2$ the value of $S_{\mathcal{K}}$ is an order of magnitude higher for the value of $\mathcal{J}_r = 1.1$ (ergodic case) when compared to $\mathcal{J}_r = 5$ (with suppressed ergodicity). This feature can also be observed by the appearance of a few distinct peaks against a vanishing background for the case when $\mathcal{J}_r = 5$, indicating that even after significantly long time-evolution the time evolved operator only has non-vanishing overlap with a small subset of all possible Krylov basis vectors, irrespective of the initial operator being $\hat{O}_1$ or $\hat{O}_2$. While the overlap with other basis states is not exactly zero, the actual values of the coefficients are vanishingly small. Comparing this for the case when $\mathcal{J}_r = 1.1$, there are several non-zero coefficients, resulting in many peaks along with relatively large values of coefficients found in between the peaks. Apart from the operators $\hat{O}_1$ and $\hat{O}_2$, which have a sparse structure, we also consider a random operator $\hat{O}_R$ defined as:
	\begin{equation}
		\hat{O}_R = \otimes_{i = 1}^{7} \hat{u}_i
	\end{equation}  
	where $\hat{u}_i$ is a single qubit random unitary operator, drawn from the set of Haar unitaries. This random operator has a much higher support in terms of the Hilbert-Schmidt basis states, and we have also calculated the value of $S_{\mathcal{K}}$ for $\hat{O}_R$ being the initial operator. We find that for $\mathcal{J}_r = 1.1$, the value of $S_{\mathcal{K}} (0.26)$ is nearly twice the value obtained for $S_{\mathcal{K}} (0.14)$ at $\mathcal{J}_r = 5.0$. This again indicates that for the regime wherein we have ergodicity breaking, an arbitrary initial operator spreads in a much smaller region of the allowed Krylov space than it would for the ergodic regime. This indeed is a feature observed in systems with Hilbert space fragmentation (HSF) \cite{Moudgalya_2022, Khemani_Fragmentation_2020}, for which the Krylov space analysis is a useful diagnostic. Based on the results for Krylov related calculations we have obtained so far, the operators $\hat{O}_1$ and $\hat{O}_2$ yield similar results in terms of coefficients $\{b_n\}$, variation of IPR with $\mathcal{J}_r$ and also the growth of $\mathcal{K}_C(t)$. A difference between the observables is obtained on examining the metric $S_{\mathcal{K}}$ in the Krylov space, where for every value of $\mathcal{J}_r$ within the ergodic regime, the final spread of operator $\hat{O}_2$ exceeds that for $\hat{O}_1$. This can be attributed to the IPR properties of the said operators wherein IPR($\hat{O}_1$) exceeds IPR($\hat{O}_2$) for all $\mathcal{J}_r$, indicating that a lower initial spread facilitates enhanced growth of $S_{\mathcal{K}}$ under time evolution.
	
	Note than an analytical approach also exists for ergodicity breaking in unitary dynamics \cite{Krylov_Budhaditya}, and the approach can be expanded for non-hermitian models as well to observe similar effects \cite{open_sahu1, open_sahu2, non_hermitian_krylov}. 
	
	\section{Relation with Entanglement Generation}
	\label{entanglement}
	\begin{figure*}
		\includegraphics[width=0.95\linewidth]{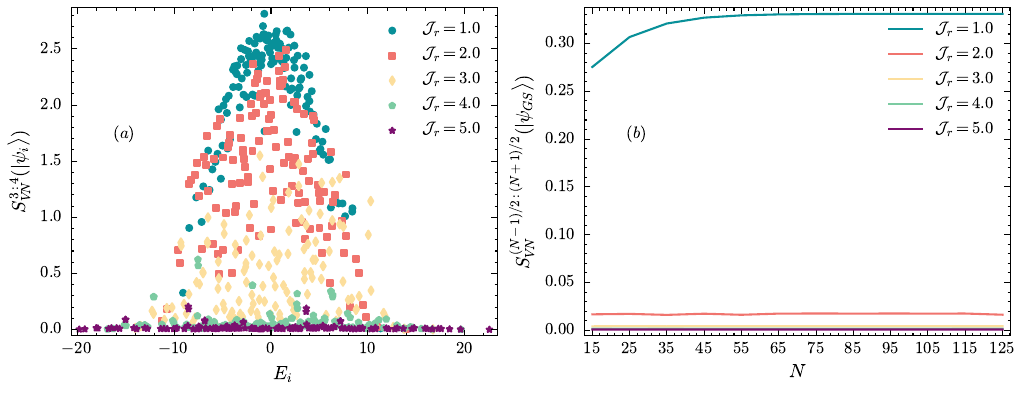}
		\caption{The entanglement values of state upon changing the parameter $\mathcal{J}_r$. In $(a)$ we have plotted the spread of all eigenstates with their energy and entanglement values represented as a point in the scatter graph for different parameters $\mathcal{J}_r$. In $(b)$, we have plotted the entanglement within ground state $\ket{\psi_{GS}}$ for different system sizes, indicating that for $\mathcal{J}_r \neq 1$ the entanglement present remains significantly diminished for even larger system sizes.}
		\label{Figure12}
	\end{figure*}
	\begin{figure*}
		\includegraphics[width=0.95\linewidth]{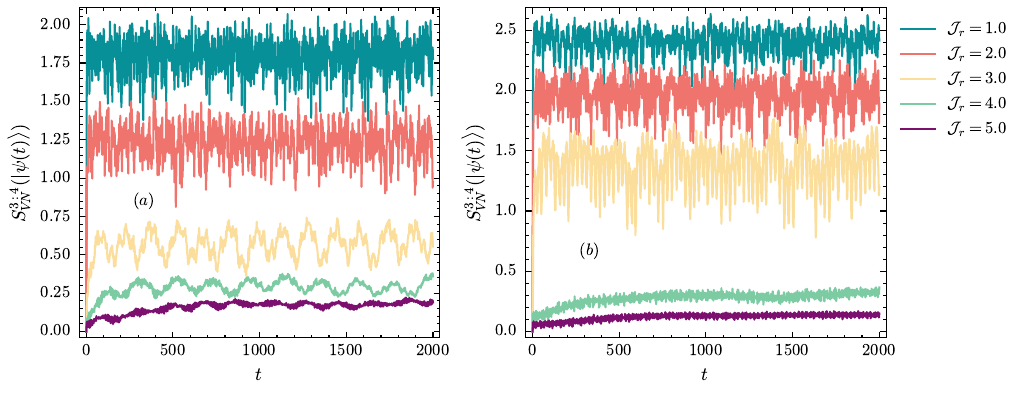}
		\caption{The growth of entanglement (across the bond between 3rd and 4th spin) with time, under unitary evolution generated by different values of $\mathcal{J}_r$ for the initial state being $(a)$ $\ket{0}^{\otimes N}$ and $(b)$ the Neel state.}
		\label{Figure13}
	\end{figure*}
	
	Before concluding with our results we shall examine the entanglement within states, and the growth of entanglement for an initial state as a final diagnostic of ergodicity breaking in the model. 
	
	In a many-body quantum model, the scaling of bipartite entanglement along with system size provides a useful diagnostic for the underlying structure of the eigenstates and thereby allows us to predict the dynamical properties of the model. A ground state of a system made up of local interaction terms with a gapped spectrum obey the area law, wherein the entanglement entropy $S_A$ of a subsystem $A$ scales with the boundary $\partial A$. This reflects the absence of long-range correlations. In contrast, when considering the excited eigenstates of a generic thermalising system the scaling relation follows the volume law given by $S_A \propto |A|$, which is consistent with ETH. In the presence of ergodicity breaking mechanisms the system is prevented from exploring the entire Hilbert space. This is accompanied by an absence of long-range correlations, and thus a departure from the volume law scaling observed in a thermalising system. In our case the entire setup consists of a single dimensional model, implying that the area law growth means that the entanglement across a particular partition for increasing system must remain constant. In the results that follow, we shall show the effects of ergodicity breaking on the eigenstates of our model and that for the system with ergodicity breaking the entanglement for ground state remains small, and significantly suppressed when compared to the non-integrable transverse field ising model. For calculating the entanglement entropy, we use the Von-Neumann measure:
	\begin{equation}
		S^{x:N-x}_{VN} = \tr(\rho_x \ln(\rho_x)) 
	\end{equation}  
	where $x:N-x$ represents the bond between $x$th and $(N-x)$th spins. Since our model consists of an odd number of spins, we take the partition $(N - 1)/2:(N + 1)/2$ for a system of size $N$. For $N = 7$, the partition we consider is between the 3rd and 4th spins, with the entanglement measure given by $S^{3:4}_{VN}$.
	
	With $N = 7$ we take different values of $\mathcal{J}_r$, and obtain the energy eigenstates for each case, plotting the energies with entanglement entropy given by $S^{3:4}_{VN}$. Thus for each value of $\mathcal{J}_r$, we get 128 data points which are shown in Fig. \ref{Figure12}$(a)$. We can see that as $\mathcal{J}_r$ reaches $5.0$, nearly all eigenstates have a near zero value of entanglement entropy. For larger system sizes as well, the entanglement entropy remains small for the ground state when $\mathcal{J}_r > 1$. For $\mathcal{J}_r = 1$ it is known from CFT results that $S_{VN}$ converges to around 0.33 which can be observed. In order to obtain these results we find out the ground state by DMRG calculations using the TeNPy library \cite{tenpy}, which can then be used for calculating entanglement. 
	
	Another metric which correlates with the transition away from ergodic dynamics is the growth of entanglement for a selected initial state under bipartition. To obtain this result, we again evaluate the entanglement entropy given by $S^{3:4}_{VN}$ for an specified initial state under time evolution, taking different values of $\mathcal{J}_r$ from $1.0$ to $5.0$ (see Fig. \ref{Figure13}). The initial states that we consider are $(a)$ the all down $\ket{0}^{\otimes N}$ state and $(b)$ the Neel state. Across both these intial states the growth of entanglement is rapid and reaches a high saturation values for ergodic dynamics, which corresponds to $\mathcal{J}_r = 1.0$. Increasing the value of $\mathcal{J}_r$ causes a systematic reduction in the saturation values of entanglement entropy, indicating that the system in region with ergodicity breaking has much lower entanglement generating capacity. These results are not surprising, as we have previously seen that the system eigenstates all have much lower entanglement with $\mathcal{J}_r$ being high. Therefore, any superposition of the eigenstates resulting out of unitary evolution is expected to show diminished values of entanglement. This also explains the similar results arising out of considering different initial states. The growth of entanglement is a well known diagnostic for ergodic dynamics, with chaotic models leading to high rates of entanglement generation. This property is irrespective of the initial state we start out with, and adds to the results presented in terms of OTOC growth and spectral properties in validating the suppression of ergodicity.
	
	\section{Summary and Conclusions}
	\label{conclusions}
	
	In our present work, we have presented a variant of the non-integrable transverse field Ising model with an added notion of inhomogenity, which leads to suppressed ergodic behaviour in the associated unitary dynamics. The departure from true ergodic nature is verified on the basis of a low saturation of the OTOC (see Fig. \ref{Figure2}) and more significantly by the transition of level spacing statistics from Wigner-Dyson to Poission (see Fig. \ref{Figure3}) upon increasing the inhomogenity parameter $\mathcal{J}_r$. In the absence of disorder, we compute the time-averaged spectral form factor and find that increasing inhomogeneity progressively diminishes the ramp, signaling a clear suppression of chaotic dynamical behaviour (see Fig. \ref{Figure4}). Having a model wherein the dynamical nature is tunable, allows us to directly explore the impact of constrained dynamics in the Krylov space. 
	For this, we calculate the spread of two selected initial operators via the IPR and their time-evolved version in the corresponding Krylov basis, generated for the Hamiltonian having a particular value of the inhomogenity parameter $\mathcal{J}_r$. We find that as changing $\mathcal{J}_r$ affects the dynamical nature of the system, it also directly affects the spread of a time evolved operator in the Krylov space, with the spread being limited for the case with constrained ergodicity. Apart from the sparse initial operators we consider, the effect is also apparent for several instances of random operators indicating the robustness of the said observation. This is in line with the expected behaviour and has also been observed in Floquet systems with dynamical freezing, which is another example of an ergodicity breaking mechanism \cite{asmi_halder_frozen_dynamics}. Recent studies have also explored effects of ergodicity breaking with the introduction of long-range hopping terms following a similar approach \cite{non_hermitian_krylov}, which is a setup that has also shown the localisation of Majorana zero modes in a hybrid Kitaev chain \cite{Rajiv_PRB}. The said qualitative link between ergodicity-breaking and transport properties maybe further explored. 
	
	Having the associated Krylov basis for different values of the inhomogeneity parameter, we also calculate the metric of Krylov complexity. We again find a strong dependence of this metric on the initial operator we start out with, similar to the results obtain by the authors of Ref. \cite{Aravinda_Sir_PRE} for the case of Floquet operators. The dynamics being continuous in time yield a more general result, even if analytically viable for a set of well-structured and sparse initial operators (see Appendix \ref{IPR_Calculation}). With this result, it becomes apparent that when analysing results for Krylov complexity and its saturation behaviour it is important to also consider properties of the initial operator. However, we also find that the initial growth phase of Krylov complexity can indeed serve as an indicator of dynamical behaviour (see Fig. \ref{Figure10}) as we find significant differences in growth rate depending on the ergodicity of dynamics, in a region wherein the operators have near similar IPR. While associated with constructing the Krylov basis vectors we find that dispersion of the Arnoldi coefficients to also serve as a reliable indicator of the underlying dynamics of the system, which shows a systematic variation with the inhomogeneity parameter $\mathcal{J}_r$. In the final section, we evaluate the bipartite entanglement of the eigenstates corresponding to the Hamiltonian under different values of inhomogeneity. We find that the associated entanglement entropy drops significantly on increasing the value of $\mathcal{J}_r$, which is a further signature of restricted ergodicity (see Fig. \ref{Figure12}). We also find that the ground state entanglement remains at a vanishingly small value for $\mathcal{J}_r \neq 1$ even with larger system sizes, obtaining the ground state and the corresponding entanglement entropy using the DMRG technique. This also implies that the entanglement generating capacity of the model becomes significantly diminished, as is expected from a model showing slow thermalisation. 
	
	In conclusion,  our work presents evidence to support our observations of restrained ergodicity in the inhomogeneous transverse field Ising model by studying the variation of several measures related to quantum dynamics. While not a generic feature, for this particular class of models it gives us an access to a system wherein the dynamical nature can be tuned by changing an externally added parameter. We believe that this observation shall enhance the usefulness of transverse field Ising model and shall allow for more investigations related to ergodicity constraining mechanisms in quantum dynamics.  
	
	\section*{Acknowledgment}
	We wish to acknowledge the resources of supercomputing facility 'Param Shivay' at IIT(BHU), which were used to generated the results presented in this work.  
	
	\section*{Data Availability}
	The data that support the findings of this article are openly available \cite{github_data_reference_inhomogeneous}.
	
	\bibliography{inhomogeneous_scrambling}
	\appendix
	\section{Spectral properties based on interaction}
\label{Spectral_properties}
\begin{figure*}
	\includegraphics[width = \textwidth]{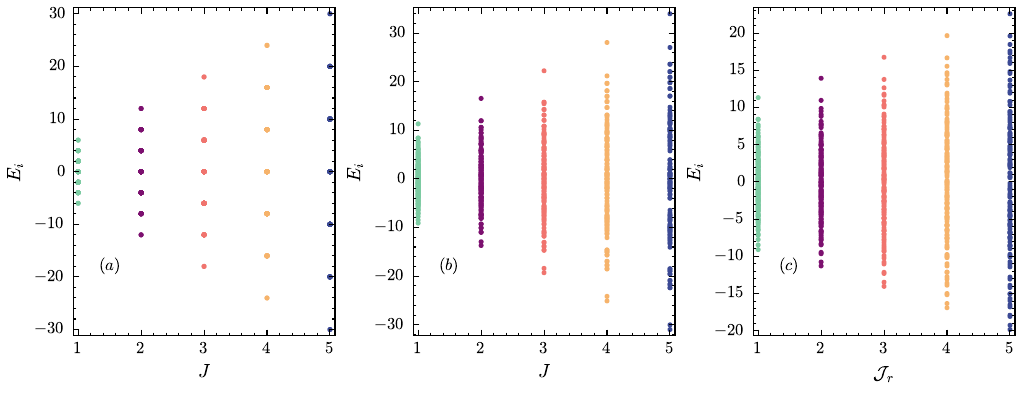}
	\caption{Variation of the eigenspectrum for transverse field Ising model Hamiltonian under the condition $(a.)$ $h_x = 0.0$, $h_z = 0.0$ with increasing values of $J$ $(b.)$ $h_x = 1.05$, $h_z = 0.5$ with increasing values of $J$ and $(c.)$ $h_x = 1.05$, $h_z = 0.5$ with increasing values of $\mathcal{J}_r$. For the figures $(a.) \& (b.)$, the value of $\mathcal{J}_r = 1.0$, indicating a uniform change of coupling across the hamiltonian, whereas in the case of $(c.)$ there is an instance of inhomogeneity affecting the energy eigenspectrum.}
	\label{Figure14}
\end{figure*}
In this section, we shall present some results depicting the effect on energy eigenspectrum coming out of changing the inhomogeneity parameter $\mathcal{J}_r$. Following the Hamiltonian given in Eq. \ref{hamiltonian}, we will explicitly plot the energy eigenvalues corresponding to different value of interaction strength $J$ for the integrable and non-integrable version of the model, with all individual interactions $J$ being equal, before showing the case with inhomogeneous $J$ for the non-integrable model. These results shall give us a basic intuition about the energy level spacings, and the manner in which selectively modifying the interaction strengths can gives us the variations observed, for example in figure Fig.\ref{Figure5}.

Changing the values of $J$ can be seen to expand the energy spectrum of the Hamiltonian (see Fig. \ref{Figure14} $(a.)$). This is because the action of spin-spin interaction has a diagonal contribution to the energy eigenvalues, and therefore increasing the term $J$ by a factor induces an increased range of the energy eigenvalues. This change is also accompanied in the non-integrable regime, however, it is important to note that the level spacing characteristics of the Hamiltonian still remain unchanged upon changing $J$ uniformly throughout the system. This is because the level spacing statistics are defined for a eigenspectrum where the difference between successive energy levels must be normalised to unity. Following this normalisation, the effect of an increased value of $J$ is removed. For example, in the results presented in Figure Fig. \ref{Figure14} $(b.)$ for the non-integrable, but homogeneous version of the Hamiltonian, while the range of energy eigenvalues are expanded, the level spacing statistics remain Wigner-Dyson which is associated with a chaotic model.

Changing the interaction strength $J$ partially within the system, like in the inhomogeneous model we define, there is a chance for partial shifting within the energy eigenvalues. This indeed has the potential to change the level spacing nature of the eigenspectrum, as we find for our model. Comparing the case when $J = 5$ and $\mathcal{J}_r = 5$, between \ref{Figure14} $(b.) \& (c.)$ corresponding to the homogeneous and inhomogeneous non-integrable setup, we can see a clear deviation from level repulsion to trivial energy spacing. Note the this transition is not general, and shall only arise in specific examples, like the one we consider. Finally, while we only discuss the case where $\mathcal{J}_r$ is greater than unity, the same deviation towards ergodicity breaking is observed when we take $\mathcal{J}_r \in (0,1)$. This indicates that for our setup $\mathcal{J}_r = 1$ is truly ergodic, with any deviation from this value gives signatures of ergodicity breaking.    

\section{Calculations for OTOC}
\label{OTOC_Calculation}
To derive the analytical variation for the OTOC in an inhomogeneous transverse field ising model we use a toy model with $N=3$ and observe the present structures. Following this, we can generalize the formula for any system size using the obtained structure.

The explicit Hamiltonian for the system size of $N=3$ can be written as follows, using Eq. \ref{hamiltonian}:
\begin{multline}
	\hat H = -J_1\sigma_1^z \sigma_2^z - J_2  \sigma_1^z \sigma_2^z - \\ h_x(\sigma_1^x + \sigma_2^x+\sigma_3^x) - h_z(\sigma_1^z + \sigma_2^z +  \sigma_3^z)
\end{multline}
The time evolution of an operator $\hat W$ can be defined by using the Baker-Campbell-Hausdorff (BCH) formula that is given as:
\begin{multline}
	\hat W(t)=\hat U \hat W(0) \hat U = \hat W(0) + it[\hat H,\hat W(0)] + \\ \frac{(it)^2}{2!}[\hat H,[\hat H,\hat W(0)]] + ....
	\label{BCH_Eq}
\end{multline}
We consider the observables at site $1$ and $n$ given as $\hat W=\sigma_1^z$ and $\hat V=\sigma_n^z$ respectively, as specified in the main text. Then OTOC is defined as,
\begin{align}
	C(t)=\tr([\hat W(t) , \hat V]^2)=\tr([\hat \sigma_1^z(t) , \hat \sigma_n^z]^2).
\end{align}
which can be expanded in the following manner, replacing $\sigma_1^z(t)$ with its time-evolved form given by equation Eq. \ref{BCH_Eq}:
\begin{multline}
	\label{A4}
	C(t) = \tr\Big([ \sigma_1^z(0) + it[H,\sigma_1^z(0)] + \\ \frac{(it)^2}{2!}[H,[H,\sigma_1^z(0)]] + ..., \sigma_n^z]^2\Big)
\end{multline}
To evaluate these commutators, we make use of the properties of Pauli operators, which are defined as:
\begin{align}
	[\sigma_i^{\alpha},\, \sigma_j^{\beta}] 
	&= 2 i\, \delta_{ij}\, \epsilon_{\alpha\beta\gamma}\, \sigma_i^{\gamma}, \\[6pt]
	(\sigma_i^{\alpha})^2 &= \mathbb{1}, \\[6pt]
	\left[ \sigma_a \otimes \sigma_c,\,
	\sigma_b \otimes \sigma_d \right]
	&= 2 i\, \epsilon_{abe}\, \delta_{cd}\,
	(\sigma_e \otimes \mathbb{1})
	+ 2 i\, \epsilon_{cdf}\, \delta_{ab}\,
	(\mathbb{1} \otimes \sigma_f),
\end{align}
where  $\epsilon_{ijk}$  is the Levi-Civita tensor.

Taking $d$ as the separation between observables involved in the OTOC expression, we have $n = 1 + d$. For $d = 0, n = 1$, it is found that of all commutator terms in equation Eq. \ref{A4}, the first order term is the most significant in characterising OTOC growth. This is because for small $t (<< 1)$, the lower-most power is the largest contributing term, and the first order commutator is the smallest power term which survives. This then commutes with $\sigma_n^z$, where $n = 1$ in this case. 

Therefore the resulting expression for $C(t)$, which involves a further squaring operation is:
\begin{equation}
	C(t) = (-t)^2 \tr([[H,\sigma_1^z(0)],\sigma_1^z]^2) + \mathcal{O}(t^3)
	\label{ct1}
\end{equation} 
The first order commutator can also be calculated easily, and is given as:
\begin{equation}
	[H,\sigma_1^z(0)] = 2ih_x \sigma_1^y,
\end{equation}
which can them be substituted in Eq. \ref{ct1}, giving the expression:
\begin{equation}
	C(t) = 16h_x^2t^2 + \mathcal{O}(t^3).
\end{equation}
Now, we present the calculation for the following higher order commutators involving $\hat{H}$, that appear in the expansion given in Eq. \ref{A4}.
\begin{multline}
	[H, [H,\sigma_1^z(0)]] = 4h_x^2 \sigma_1^z - 4h_xh_z\sigma_1^x - \\ 4J_1h_x\sigma_1^x\sigma_2^z,
\end{multline}
and proceeding iteratively, we have:
\begin{multline}
	[H,[H, [H,\sigma_1^z(0)]]] = 8ih_x(J_1^2 + h_x^2 + h_z^2)\sigma_1^y \\ - 8iJ_1h_x^2\sigma_1^x\sigma_2^y + 16iJ_1h_xh_z\sigma_1^y\sigma_2^z.
	\label{3oc}
\end{multline}
Note that for $d = 1$, i.e. for operator $\hat{V}$ being $\sigma_2^z$, the lowest term which does not vanish is the third order commutator, $[H,[H, [H,\sigma_1^z(0)]]]$ specified above. Following the same reasoning that lead us to Eq. \ref{ct1} we can write an expression for $C(t)$ as:
\begin{equation}
	C(t) = k \cdot \frac{t^6}{(3!)^2} + \mathcal{O}(t^7),
\end{equation}  
with $k$ being a constant. This generalisation holds for higher values of $d$, wherein the lowest order term that does not vanish is the $(2d+1)th$ order commutator. From this variation, we have our result that for a given $d$, assuming a transverse field ising model underneath, the expression for OTOC follows a power law given as:
\begin{equation}
	C(d,t) \propto \frac{t^{2(2d+1)}}{((2d+1)!)^2}. 
\end{equation}
Further, it can be seen from Eq. \ref{3oc} that for $d = 1$, there is no role played by the inhomogeneity factor $e$. That only appears when $d \ge (N-1)/2$, or what would be the fifth order commutator in our case.

\section{Dependence of $\mathcal{K}_C(t)$ on initial operators}
\label{IPR_Calculation}

In this section we shall attempt to find the dependence of long-time behaviour for Krylov complexity $\mathcal{K}_C(t)$ on the initial operator. For this calculation we shall make use of the energy eigenbasis to represent the initial operator and its time evolved version. We shall then obtain the corresponding coefficients for the general Krylov basis vectors. We start with the initial operator $\hat{O}$ at $t = 0$. This can be given as:
\begin{equation}
	\hat{O} = \sum_{\alpha,\beta} O_{\alpha.\beta}\ket{E_{\alpha}}\bra{E_{\beta}},
\end{equation} 
following a decomposition in terms of the projector operators made up of the energy eigenstate $\ket{E_i}$. Thus, at time $t$ the operator is given as:
\begin{equation}
	\hat{O}(t) = e^{-i\hat{H}t/\hbar}\cdot\hat{O}\cdot e^{i\hat{H}t/\hbar}.
	\label{time_evolved_op}
\end{equation}   
The unitary operator takes a diagonal form in the energy eigenbasis, which is given as:
\begin{equation}
	e^{-i\hat{H}t} = \sum_{\gamma} exp(-iE_{\gamma}t/\hbar)\ket{E_{\gamma}}\bra{E_{\gamma}}.
\end{equation} 
Substituting this representation in the above equation (Eq. \ref{time_evolved_op}), we have:
\begin{equation}
	\hat{O}(t) = \sum_{\alpha, \beta} exp(i(E_{\beta} - E_{\alpha})t)O_{\alpha,\beta}\ket{E_{\alpha}}\bra{E_{\beta}}. 
\end{equation} 
Here, we have taken $\hbar = 1$. Further, the term $\ket{E_{\alpha}}\bra{E_{\beta}}$ is a projector operator, and maybe represented as $\hat{P}_{\alpha,\beta}$. Thus, the above expression can be modified as follows:
\begin{equation}
	\hat{O}(t) = \sum_{\alpha,\beta} exp(-i(E_{\alpha} - E_{\beta})t) O_{\alpha,\beta}\hat{P}_{\alpha,\beta}.
	\label{operatorfinal}
\end{equation}
Now, the same operator $\hat{O}(t)$ can also be expressed in terms of the Krylov basis states as:
\begin{equation}
	\hat{O} = \sum_n \phi_n(t)\ket{\mathcal{W}_n},
	\label{krylov_basis}
\end{equation}
with $\ket{K_n}$ being the operator space Krylov basis vector. Using the above expression in Eq. \ref{operatorfinal} and Eq. \ref{krylov_basis}, we have:
\begin{equation}
	\phi_n(t) = \sum_{\alpha,\beta} exp(-i(E_{\alpha} - E_{\beta})t)O_{\alpha,\beta}\langle \mathcal{W}_n|\hat{P}_{\alpha,\beta}\rangle.
\end{equation}
Using this relation, we can now find an expression for $|\phi_n(t)|^2 = \phi^*_n(t)\phi_n(t)$. This is given as:
\begin{eqnarray}
	|\phi_n(t)|^2 = \sum_{\alpha,\beta,\gamma,\delta} \Biggr[ exp(i(E_{\beta} - E_{\alpha} - E_{\gamma} + E_{\delta})t) O_{\alpha,\beta} O^{*}_{\gamma,\delta} \nonumber \\ \langle \mathcal{W}_n|\hat{P}_{\alpha,\beta} \rangle \langle  \mathcal{W}_n|\hat{P}_{\gamma,\delta} \rangle \Biggr] \,\,\,\,\,\,\,\,\,\,\
\end{eqnarray}  
We now consider the long time average of the term $|\phi_n(t)|^2$, which is given as:
\begin{equation}
	|\bar{\phi}_n|^2 = \frac{1}{T}\int_{0}^{T} |\phi_n(t)|^2 dt; \,\,\,\, T \rightarrow \infty.
\end{equation}
In this integral, it can be assumed that upon adding several terms of $|\phi_n(t)|^2$ at different values of $t$ leads to the off-diagonal terms canceling each other out due to the appearance of random phases, as explained in reference \cite{Krylov_Standard_Paper}. This leads to the following expression:
\begin{equation}
	|\bar{\phi}_n|^2 = \sum_{\alpha,\beta} |O_{\alpha,\beta}|^2 \big( \langle \mathcal{W}_n | \hat{P_{\alpha,\beta}} \rangle \big)^2.
\end{equation}
The above expression also has associated constraints such as:
\begin{equation}
	\sum_n |\bar{\phi}_n|^2 = 1 \,\,\,\,\, \text{and} \,\,\,\,\, \sum_{\alpha,\beta} |O_{\alpha,\beta}|^2 = 1
	\label{constraints}
\end{equation}
These constraints arise due to being the normalisation process we carry out for the operator $\hat{O}$ before generating the Krylov basis vectors. Thus, we can now have the time averaged value of the Krylov complexity $\mathcal{K}_C(t)$. This is given as:
\begin{equation}
	\bar{\mathcal{K}}_C = \sum_{n} n |\bar{\phi}_n|^2
\end{equation}
Now, substituting the expression for $|\bar{\phi}_n|$ we have:
\begin{eqnarray}
	\bar{\mathcal{K}}_C = \sum_{n} n \Bigg[ \sum_{\alpha,\beta} |O_{\alpha,\beta}|^2 \big( \langle \mathcal{W}_n | \hat{P}_{\alpha,\beta} \rangle \big)^2 \Bigg] \\
	= \sum_{\alpha,\beta} \Bigg[ \sum_n n\big( \langle \mathcal{W}_n | \hat{P}_{\alpha,\beta} \rangle \big)^2 \Bigg]|O_{\alpha,\beta}|^2
\end{eqnarray}
Given the variation of $\mathcal{K}_C(t)$ with time and its saturation in the long time limit, $\bar{\mathcal{K}}_C$ can be treated as a direct indicator of the saturation value of $\mathcal{K}_C(t)$ at $t \rightarrow \infty$. 

In our results we obtain two distinct kinds of variations for the saturation value of $\mathcal{K}_C(t)$. For the ergodic regime $\mathcal{K}_C(t)$ decreases on the increase of initial operator $\hat{O}$. For the regime with ergodicity breaking, this trend is reversed and increasing the IPR increases $\mathcal{K}_C(t)$. Both of these observed behaviours can be explained using the expression for $\bar{\mathcal{K}}_C$ obtained above and the constraints given in Eq. \ref{constraints}. Although not generic, the approach we follow explains our findings for the sparse operators $\hat{O}_1$ and $\hat{O}_2$. 

We can separate the previous expression obtained for $\bar{\mathcal{K}}_C$ between the diagonal and off-diagonal parts of the inital operator as follows:
\begin{eqnarray}
	\bar{\mathcal{K}}_C = \sum_{\alpha} \Bigg[ \sum_n n\big( \langle \mathcal{W}_n | \hat{P}_{\alpha,\alpha} \rangle \big)^2 \Bigg]|O_{\alpha,\alpha}|^2 + \nonumber \\  \sum_{\substack{\alpha,\, \beta \\ \alpha \ne \beta}} \Bigg[ \sum_n n\big( \langle \mathcal{K}_n | \hat{P}_{\alpha,\beta} \rangle \big)^2 \Bigg]|O_{\alpha,\beta}|^2
\end{eqnarray}
The expression for IPR is given as: $IPR(\hat{O}) = \sum_{\alpha} |O_{\alpha,\alpha}|^2$. Since the operator is normalised, it implies $\sum_{\alpha,\beta} |O_{\alpha,\beta}|^2 = 1$. Thus, a decrease in IPR increases the contribution of off-diagonal terms $|O_{\alpha,\beta}|^2, (\alpha \neq \beta)$, which increases the weight of summation $n \cdot \sum_{\alpha,\beta} (\langle \mathcal{W}_n | \hat{P}_{\alpha,\beta} \rangle)^2$ that involves only positive terms. This contributes to the higher values of $\bar{\mathcal{K}}_C$ and thereby predicts a higher saturation value of $\mathcal{K}_C(t)$. This calculation explains the observed nature of $\mathcal{K}_C(t)$ in the ergodic regime, which is associated a high value of $S_{\mathcal{K}}$. This implies that almost all of $\phi_n(t)$ are non-zero, and in this regime decreasing IPR enhances $\bar{\mathcal{K}}_C$.

On the other hand, for the region with constrained ergodicity, increasing the inhomogeneity parameter $\mathcal{J}_r$, increases the IPR for operator $\hat{O}_1$ and $\hat{O}_2$ (see Fig. \ref{Figure8}), but also decreases $S_{\mathcal{K}}$ (see Fig. \ref{Figure11}). This implies fewer non-zero terms in the summation $\sum_n |\phi_n|^2 = 1$, which automatically increases the numerical values of the non-zero terms in the summation to compensate. This thereby increases the values of $\sum_n n|\phi_n(t)|^2$. Thus, in the region with broken ergodicity, for the sparse operators $\hat{O}_1$ and $\hat{O}_2$, an increase in IPR is associated with fewer number of terms having finite contribution in the set $\{ \langle \mathcal{W}_n | \hat{O}(t) \rangle \}_n$. Due to the requirement of normalisation the remaining terms are enhanced, and this leads to an increase in the average Krylov complexity $\bar{\mathcal{K}}_C$ and by extension the saturation value of $\mathcal{K}_C(t)$.

\end{document}